# Harmonic-Gaussian double-well potential stochastic resonance with its application to enhance weak fault characteristics of machinery

Zijian Qiao[1,2,3,4], *, Shuai Chen[1], Zhihui Lai[5], Shengtong Zhou[2], Miguel A. F. Sanjuán[6]

1 School of Mechanical Engineering and Mechanics, Ningbo University, Ningbo 315211, China

2. State Key Laboratory of Performance Monitoring and Protecting of Rail Transit Infrastructure, East China Jiaotong University, Nanchang 330013, China

3. Yangjiang Research Center Advanced Energy Science and Technology Guangdong Laboratory (Yangjiang Offshore Wind Power Laboratory), Yangjiang 529500, China

4. Zhejiang Provincial Key Laboratory of Part Rolling Technology, Ningbo 315211, China

5. Guangdong Provincial Key Laboratory of Micro/Nano Optomechatronics Engineering, College of Mechatronics and Control Engineering, Shenzhen University, Shenzhen 518060, China

6. Nonlinear Dynamics, Chaos and Complex Systems Group, Departamento de Física, Universidad Rey Juan Carlos, Tulipán s/n, Móstoles, 28933, Madrid, Spain

**Abstract:**

Noise is ubiquitous and unwanted in detecting weak signals, which would give rise to incorrect filtering frequency-band selection in signal filtering-based methods including fast kurtogram, teager energy operators and wavelet packet transform filters and meanwhile would result in incorrect selection of useful components and even mode mixing, end effects and etc. in signal decomposition-based methods including empirical mode decomposition, singular value decomposition and local mean decomposition. On the contrary, noise in stochastic resonance (SR) is beneficial to enhance weak signals of interest embedded in signals with strong background noise. Taking into account that nonlinear systems are crucial ingredients to activate the SR, here we investigate the SR in the cases of overdamped and underdamped harmonic-Gaussian double-well potential systems subjected to noise and a periodic signal. We derive and measure the analytic expression of the output signal-to-noise ratio (SNR) and the steady-state probability density (SPD) function under approximate adiabatic conditions. When the harmonic-Gaussian double-well potential loses its stability, we can observe the antiresonance phenomenon, whereas adding the damped factor into the overdamped system can change the stability of the harmonic-Gaussian double-well potential, resulting that the antiresonance behavior disappears in the underdamped system. Then, we use the overdamped and underdamped harmonic-Gaussian double-well potential SR to enhance weak useful characteristics for diagnosing incipient rotating machinery failures. Theoretical and experimental results show that adjusting both noise intensity and system parameters can activate overdamped and underdamped harmonic-Gaussian double-well potential SR in which there is a bell-shaped peak for the SNR. Additionally, the underdamped harmonic-Gaussian double-well potential SR is independent of frequency-shifted and rescaling transform to process large machine parameter signals and outperforms the overdamped one. Finally, comparing the advanced robust local mean decomposition (RLMD) method based on signal decomposition and the wavelet transform method based on noise cancellation or infogram method based on signal filtering, the overdamped or underdamped harmonic-Gaussian double-well potential SR methods characterize a better performance to detect a weak signal. Fault characteristics in the early stage of failures are successful in improving the incipient fault characteristic identification of rolling element bearings.

---

*Corresponding author.
E-mail address: zijianqiao@hotmail.com, qiaozijian@nbu.edu.cn (Z. Qiao).

## 1. Introduction

Noise is ubiquitous but unwanted in detecting weak signals [1], but noise in biological systems can be used to amplify weak signals embedded by a strong background noise. Such an ingenious phenomenon is observed in a bistable nonlinear system, namely stochastic resonance (SR) [2]. SR is a kind of synchronization mechanism among the nonlinear systems, noise and a weak periodic signal, which takes place to activate the SR for amplifying weak useful signals [3].

SR has been investigated from theory to engineering application widely [4-6]. Among three ingredients for activating SR including noise, nonlinear systems and weak useful signals, nonlinear systems are crucial ingredients for extracting weak useful signals and moreover can harvest the energy of noise located at the whole frequency band of a noisy signal to enhance or amplify a weak useful signal. For this purpose, most of scholars pay attention to exploring the behaviors of SR in novel nonlinear systems from bistable [7] to multistable ones [8-10], from overdamped [11] and underdamped [12] to fractional-order [13] ones, and even from cascaded [14] and coupled [15, 16] to time-delayed feedback [17] ones and biological systems [18, 19].

Because the bistable system is most classical among them, it has been investigated, such as classical bistable potential overdamped systems, noisy confined bistable potential overdamped systems [20], asymmetric bistable potential overdamped systems [21], classical bistable potential underdamped systems, noisy bistable potential fractional-order systems [22] and E-exponential potential underdamped systems [23, 24]. The E-exponential potential named by the references [23, 24] is a narrow version of the harmonic-Gaussian double-well potential. The references above show that overdamped and underdamped harmonic-Gaussian double-well potential SR has not been studied systematically in theory and further applied to enhance incipient fault identification of machinery for providing a tutorial of other readers and researchers on the SR in the overdamped and underdamped systems with novel generalized double-well potentials yet. Even, the comparison between overdamped and underdamped harmonic-Gaussian double-well potential SR has not been made in theory and engineering applications. Therefore, this paper attempts to investigate the SR in the overdamped and underdamped harmonic-Gaussian double-well potential systems theoretically and then apply it to enhance weak fault characteristics and diagnose incipient faults of machinery. Additionally, some comparisons with other advanced signal processing techniques including signal decomposition-based and noise cancellation or signal filtering-based methods for enhancing weak fault characteristics of machinery are given.

The remainder of this paper is organized as follows. Section 2 and Section 3 investigate the overdamped and underdamped harmonic-Gaussian double-well potential SR by deriving the analytic expressions of signal-to-noise ratio (SNR) and steady-state probability density (SPD) functions, respectively. In Section 4, we apply the overdamped and underdamped harmonic-Gaussian double-well potential SR to enhance weak fault characteristics and incipient fault identification of rolling element bearings. Finally, conclusions are drawn in Section 5.

## 2. Overdamped harmonic-Gaussian double-well potential SR

The overdamped Langevin equation driven by a harmonic-Gaussian double-well potential under the action of random noise and a periodic signal can be described as [25]

$$\frac{dy}{dx} = -\frac{\partial U(x)}{\partial x} + A\cos(\omega_0 t) + \varepsilon(t) \tag{1}$$

where $A$ and $\omega_0$ are the amplitude and angular frequency of the periodic signal respectively, and $\varepsilon(t)$ is the Gaussian white noise with mean zero and variance $D$ i.e. noise intensity.

The harmonic-Gaussian double-well potential which is a variant of a double-well potential can be expressed as [26]

$$U(x) = \frac{k}{2}x^2 + \alpha\exp(-\beta x^2) \tag{2}$$

where two stable states and one unstable state are located at $x_\pm = \pm\sqrt{\ln(2\alpha\beta/k)/\beta}$ and $x_u = 0$ respectively, and the barrier height is $\Delta U = \alpha - k[1 + \ln(2\alpha\beta/k)]/(2\beta)$. To ensure the stability of the harmonic-Gaussian double-well potential, this condition $\ln(2\alpha\beta/k) > 0$ must be satisfied, further $k < 2\alpha\beta$. When $k = 1$, Fig. 1(a) shows the harmonic-Gaussian double-well potential under different system parameter sets $(\alpha, \beta)$, while Fig. 1(b) depicts those with varying $k$. It is seen from Fig. 1(a) that adjusting the system parameter $\beta$ controls the potential-well width whereas the potential-barrier height nearly keeps unchanged, but varying $\alpha$ changes the potential-barrier height whereas the potential-well width nearly remains unchanged. Such a behavior is helpful to tune the potential-well width and depth individually to activate the optimal harmonic-Gaussian double-well potential SR. Meanwhile, it is found from Fig. 1(b) that adjusting $k$ can also change the slope of the harmonic-Gaussian double-well potential.

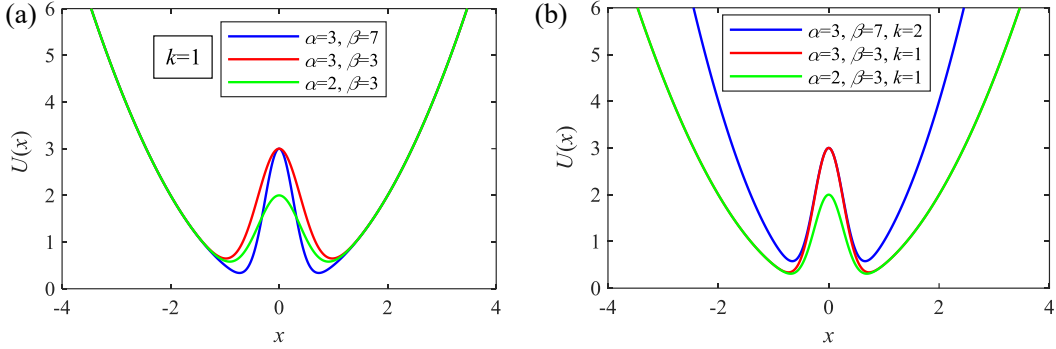

**Fig. 1** Harmonic-Gaussian double-well potentials under different parameter sets (a) $(\alpha, \beta)$ and (b) $(\alpha, \beta, k)$.

The Langevin equation in Eq. (1) can be transformed as further [27]

$$\frac{\partial \rho(x,t)}{\partial t} = -\frac{\partial}{\partial x}[-kx + 2\alpha\beta x\exp(-\beta x^2) + A\cos(\omega_0 t)]\rho(x,t) + D\frac{\partial^2}{\partial x^2}\rho(x,t) \quad (3)$$

where $\rho(x,t)$ is the probability density function (PDF) of the stochastic process $x(t)$ which denotes the transition trajectory of Brownian particles in the harmonic-Gaussian double-well potential as time varies. The corresponding SPD function can be denoted as

$$\rho_s(x,t) = \frac{N(t)}{\sqrt{D}}\exp\left[-\frac{\emptyset(x,t)}{D}\right] \quad (4)$$

where $N(t)$ is the normalization constant and $N(t) = \sqrt{D}/\int_{-\infty}^{\infty}\exp[-\emptyset(x,t)/D]dx$, and $\emptyset(x,t)$ is the generalized potential

$$\emptyset(x,t) = U(x) - xA\cos(\omega_0 t) \ . \quad (5)$$

Assuming that the periodic signal $A\cos(\omega_0 t)$ can satisfy the requirement of small parameters under approximate adiabatic conditions, i.e., $\omega_0$ is larger than the characteristic relaxation time in double potential wells [28]. Then, the transition rates between the two stable states are given by the Kramers-like formulas [29]

$$W_{\pm}(x,t) = \frac{\sqrt{|U''(x_{\pm},t)U''(x_u,t)|}}{2\pi} \exp\left[\frac{\emptyset(x_{\pm},t)-\emptyset(x_u,t)}{D}\right] \quad (6)$$

where the notation $|\cdot|$ denotes the absolute value and

$$U''(x,t) = k - 2\alpha\beta\exp(-\beta x^2)(1 - 2\beta x^2)$$

$$U''(x_u,t) = k - 2\alpha\beta$$

$$U''(x_{\pm},t) = 2k\ln\left(\frac{2\alpha\beta}{k}\right) \quad (7)$$

$$\emptyset(x_u,t) = U(x_u,t) - x_u A\cos(\omega_0 t) = \alpha$$

$$\emptyset(x_{\pm},t) = U(x_{\pm},t) - x_{\pm}A\cos(\omega_0 t) = \frac{k}{2\beta}\left(1 + \ln\frac{2\alpha\beta}{k}\right) \mp A\cos(\omega_0 t)\sqrt{\frac{\ln(2\alpha\beta/k)}{\beta}}$$

When we introduce Eq. (7) into Eq. (6), we can obtain

$$W_{\pm}(x,t) = \frac{\sqrt{k(2\alpha\beta - k)\ln(2\alpha\beta/k)}}{\sqrt{2}\pi}$$

$$\times \exp\left[-\frac{\alpha}{D} + \frac{k(1+\ln(2\alpha\beta/k))}{2\beta D} \mp A\cos(\omega_0 t)\sqrt{\frac{\ln(2\alpha\beta/k)}{\beta D^2}}\right] \quad (8)$$

Furthermore, Eq. (8) can be transformed as

$$W_{\pm}(x,t) = f(\mu \pm \eta_0 \cos(\omega_0 t)) \quad (9)$$

where

$$\mu = \frac{\alpha}{D} - \frac{k}{2\beta D}\left(1 + \ln\frac{2\alpha\beta}{k}\right) \quad (10)$$

$$\eta_0 = \frac{A}{D}\sqrt{\frac{\ln(2\alpha\beta/k)}{\beta}} \quad (11)$$

Thus, we can simplify Eq. (8) as

$$W_{\pm}(x,t) = f(\mu \pm \eta_0 \cos(\omega_0 t)) = \frac{\sqrt{2k(2\alpha\beta-k)\ln\frac{2\alpha\beta}{k}}}{2\pi}\exp[-(\mu \pm \eta_0\cos(\omega_0 t))] \quad (12)$$

The response of the nonlinear system in Eq. (1) can be quantified using a classical measure, i.e., SNR [30]. To derive its analytic expression, the power spectral density of the system response can be described as

$$S(\Omega) = \left[1 - \frac{\alpha_1^2 \eta_0^2}{2(\alpha_0^2+\omega_0^2)}\right]\left(\frac{4c^2\alpha_0}{\alpha_0^2+\omega_0^2}\right) + \frac{\pi c^2 \eta_0^2 \alpha_1^2}{\alpha_0^2+\Omega^2}[\delta(\Omega - \omega_0) + \delta(\Omega + \omega_0)] \quad (13)$$

where

$$c = \sqrt{\frac{\ln(2\alpha\beta/k)}{\beta}} \quad (14)$$

$$\alpha_1 = \alpha_0 = \frac{\sqrt{2k\ln(2\alpha\beta/k)(2\alpha\beta-k)}}{\pi}\exp(-\mu) \tag{15}$$

Finally, the output SNR of the response of the overdamped harmonic-Gaussian double-well potential system can be derived as

$$\text{SNR} = \frac{\pi c^2 \alpha_1^2 \eta_0^2}{\alpha_0^2+\Omega^2}\Big|_{\Omega=\omega_0} \times \frac{\alpha_0^2+\omega_0^2}{4c^2\alpha_0}\left[1-\frac{\alpha_1^2\eta_0^2}{2(\alpha_0^2+\omega_0^2)}\right]^{-1} = \frac{\pi\alpha_1\eta_0^2}{4}\left[1-\frac{\alpha_1^2\eta_0^2}{2(\alpha_0^2+\omega_0^2)}\right]^{-1} \tag{16}$$

Therefore, we can analyze the function between the output SNR and system parameters using the analytic expression in Eq. (16). Figure 2 shows the output SNR of overdamped harmonic-Gaussian double-well potential SR as system parameters and noise intensity vary. It can be seen from Fig. 2(a) that the output SNR is a nonmonotonic function of noise intensity $D$ under different $k$ and the peak value of output SNR increases when $k$ raises, suggesting that adjusting $k$ is able to activate the SR in the overdamped harmonic-Gaussian double-well potential system for improving the output SNR. Similarly, adjusting $\alpha$ and $\beta$ can also maximize the output SNR, and the peak value of the output SNR declines as $\alpha$ or $\beta$ increases but the resonant noise intensity at the peak value becomes larger, as shown in Fig. 2(b) and Fig. 2(c), respectively. We visualize the two-dimensional function among SNR and two of system parameters $(\alpha, \beta, k)$, as shown in Fig. 2(d), Fig. 2(c) and Fig. 2(d). One can observe from Fig. 2(d) that a moderate parameter set $(\alpha, k)$ can improve the SNR of a given signal, whereas there exists a negative output SNR because the harmonic-Gaussian double-well potential loses its stability when $k \geq 2\alpha\beta$, resulting in an antiresonance phenomenon. Meanwhile, we fix $k$ to express the output SNR as a function of $(\alpha, \beta)$ in Fig. 2(e), indicating that only an optimal matching between $\alpha$ and $\beta$ can activate the overdamped harmonic-Gaussian double-well potential SR to enhance the weak periodic signal embedded by a strong background noise. Similarly, Fig. 2(f) also demonstrates that such a parameter matching is necessary to activate the overdamped harmonic-Gaussian double-well potential SR. When $k \geq 2\alpha\beta$, one can also see the antiresonance from Fig. 2(e) and 2(f), respectively. The above results demonstrate that the optimal parameter matching among $k$, $\alpha$ and $\beta$ is able to maximize the SR.

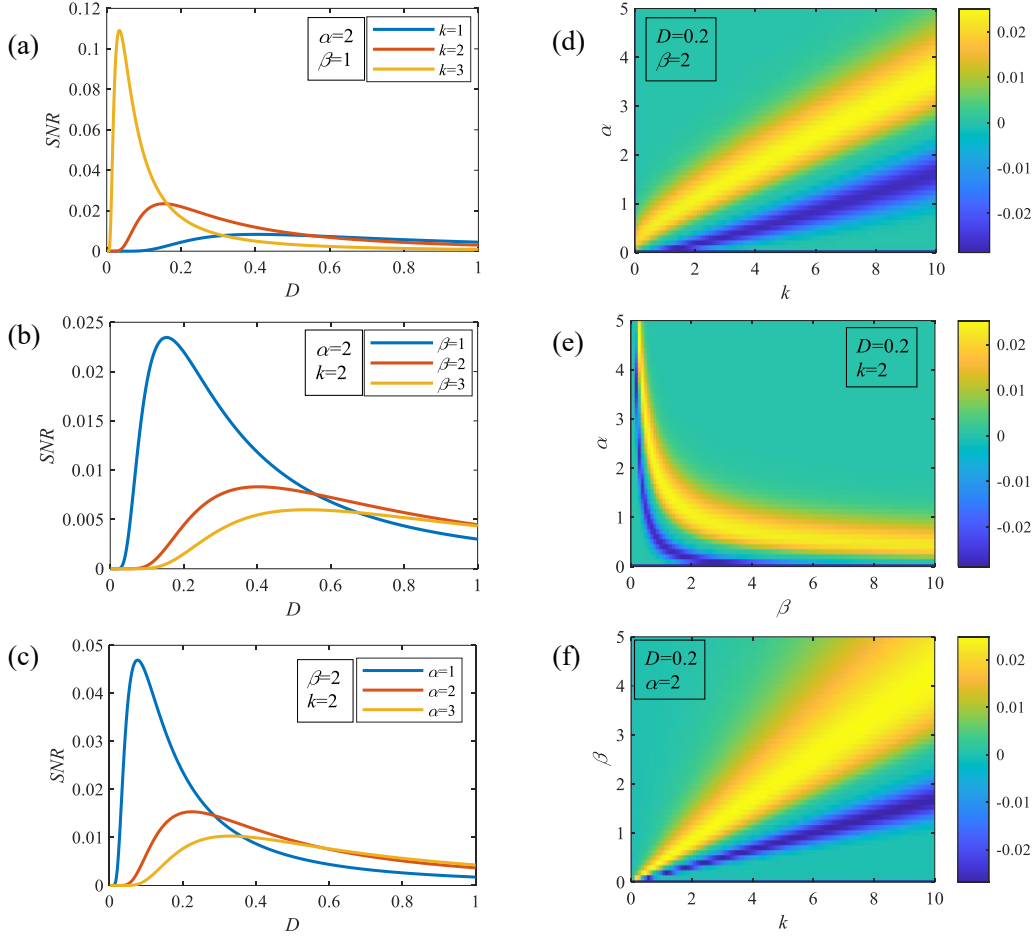

**Fig. 2** SNR of overdamped harmonic-Gaussian double-well potential SR varies with system parameters and noise intensity: SNR as a function of noise intensity under different $k$ in (a), $\beta$ in (b) and $\alpha$ in (c); SNR as a two-dimensional function of $(k,\alpha)$ in (d), $(\beta,\alpha)$ in (e) and $(k,\beta)$ in (f).

Figure 3 depicts the SPD function and the corresponding system responses. The SPD indicates the probability of Brownian particles to reside in double potential wells. It is found from Fig. 3(a) that when $D=0.3$ the particles oscillate at the right potential well located at $x_+=\sqrt{\ln(2\alpha\beta/k)/\beta}$ for activating intra-well SR, which is demonstrated by the system response in Fig. 3(b) further. When we increase the noise intensity $D$, the particles can jump across the potential barrier to go back and forth in double wells for activating the inter-well SR marked in red in Fig. 3(a), whose system response characterizes the eye-catching period marked in red in Fig. 3(b). When the noise intensity is fixed as $D=3$, two peaks of SPD decline and the corresponding system response marked in green becomes noisy.

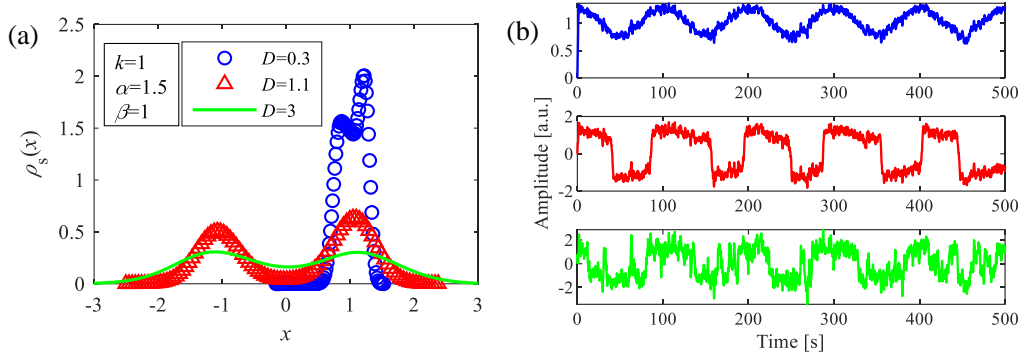

**Fig. 3** SPD functions and the corresponding system responses of overdamped harmonic-Gaussian double-well potential SR under different noise intensity: (a) the SPD functions and (b) the corresponding system responses.

## 3. Underdamped harmonic-Gaussian double-well potential SR

The underdamped harmonic-Gaussian double-well potential system subjected to a periodic signal and noise can be described as [31]

$$\frac{d^2x}{dt^2} + \gamma \frac{dx}{dt} = -\frac{\partial U(x)}{\partial x} + A\cos(\omega_0 t) + \xi(t) \tag{17}$$

where $\gamma$ is the damped factor and $\gamma > 0$. Equation (17) can be transformed as [32]

$$\begin{cases} \frac{dx}{dt} = y \\ \frac{dy}{dt} = -\gamma y - kx + 2\alpha\beta x\exp(-\beta x^2) + A\cos(\omega_0 t) + \xi(t) \end{cases} \tag{18}$$

Supposing that $A = 0$, $D = 0$, $dx/dt = 0$ and $dx/dt = 0$, we can obtain three singular points

$$\begin{pmatrix} x_\pm \\ y_\pm \end{pmatrix} = \begin{pmatrix} \pm\sqrt{\frac{\ln(2\alpha\beta/k)}{\beta}} \\ 0 \end{pmatrix}, \begin{pmatrix} x_u \\ y_u \end{pmatrix} = \begin{pmatrix} 0 \\ 0 \end{pmatrix} \tag{19}$$

Let $\partial U(x,y)/\partial x$ and $\partial U(x,y)/\partial y$ mark as $U_x(x,y)$ and $U_y(x,y)$ respectively, and then Eq. (18) can be rewritten as

$$\begin{cases} U_x(x,y) = y \\ U_y(x,y) = -\gamma y - kx + 2\alpha\beta x\exp(-\beta x^2) + A\cos(\omega_0 t) + \xi(t) \end{cases} \tag{20}$$

The linearization matrix of Eq. (18) can be calculated as

$$\begin{pmatrix} U_{xx}(x,y) & U_{xy}(x,y) \\ U_{yx}(x,y) & U_{yy}(x,y) \end{pmatrix} = \begin{pmatrix} 0 & 1 \\ -k + 2\alpha\beta\exp(-\beta x^2)[1 - 2\beta x^2\exp(-\beta x^2)] & -\gamma \end{pmatrix}$$

(21)

Further, the linearization matrix at the singular points $\left(\pm\sqrt{\ln(2\alpha\beta/k)/\beta},0\right)$ can be denoted as

$$\begin{pmatrix} U_{xx}(x,y) & U_{xy}(x,y) \\ U_{yx}(x,y) & U_{yy}(x,y) \end{pmatrix}\bigg|_{x=\pm\sqrt{\frac{\ln(2\alpha\beta/k)}{\beta}},y=0} = \begin{pmatrix} 0 & 1 \\ -\frac{k^2}{\alpha\beta}\ln\left(\frac{2\alpha\beta}{k}\right) & -\gamma \end{pmatrix} \quad (22)$$

By solving Eq. (22), the corresponding eigenvalues are calculated as

$$\beta_{1,2} = \frac{-\gamma \pm \sqrt{\gamma^2 - \frac{4k^2}{\alpha\beta}\ln\left(\frac{2\alpha\beta}{k}\right)}}{2} \quad (23)$$

Similarly, the linearization matrix at the singular point $(0,0)$ is

$$\begin{pmatrix} U_{xx}(x,y) & U_{xy}(x,y) \\ U_{yx}(x,y) & U_{yy}(x,y) \end{pmatrix}\bigg|_{x=0,y=0} = \begin{pmatrix} 0 & 1 \\ -k+2\alpha\beta & -\gamma \end{pmatrix} \quad (24)$$

The corresponding eigenvalues to the linearization matrix in Eq. (24) are

$$\lambda_{1,2} = \frac{-\gamma \pm \sqrt{\gamma^2 + 4(2\alpha\beta - k)}}{2} \quad (25)$$

Assuming that $\rho(x,y,t)$ is the PDF of the stochastic process in Eq. (18), the corresponding the Fokker-Planck equation is [33]

$$\frac{\partial \rho(x,y,t)}{\partial t} = -\frac{\partial y}{\partial x}\rho(x,y,t) - \frac{\partial}{\partial y}[-\gamma y - kx + 2\alpha\beta x\exp(-\beta x^2) +$$

$$A\cos(\omega_0 t)]\rho(x,y,t) + \gamma D \frac{\partial^2}{\partial y^2}\rho(x,y,t) \quad (26)$$

Then, the corresponding SPD function to Eq. (18) can be denoted as

$$\rho_s(x,y,t) = N(t)\exp\left[-\frac{1}{D}\left(\frac{1}{2}y^2 + \frac{k}{2}x^2 + \alpha\exp(-\beta x^2) - xA\cos(\omega_0 t)\right)\right] \quad (27)$$

where $N(t)$ stands for the normalization constant [34]

$$N(t) = \frac{1}{\int_{-\infty}^{\infty}\int_{-\infty}^{\infty}\exp\left[-\frac{\hat{U}(x,y,t)}{D}\right]dxdy} \quad (28)$$

in which $\hat{U}(x,y,t)$ denotes the generalized potential

$$\hat{U}(x,y,t) = \frac{1}{2}y^2 + \frac{k}{2}x^2 + \alpha\exp(-\beta x^2) - xA\cos(\omega_0 t) \quad (29)$$

The transition rates of particles at the singular points $(x_\pm, y_\pm) = \left(\pm\sqrt{\ln(2\alpha\beta/k)/\beta},0\right)$ can be calculated as [35]

$$W_\pm(t) = \frac{\sqrt{\beta_1\beta_2}}{2\pi}\sqrt{-\frac{\lambda_1}{\lambda_2}}\exp\left[\frac{1}{D}\left(\frac{k}{2\beta}\left(1+\ln\left(\frac{2\alpha\beta}{k}\right)\right) - \alpha \mp \sqrt{\frac{\ln(2\alpha\beta/k)}{\beta}}A\cos(\omega_0 t)\right)\right]$$

(30)

Finally, the analytic expression of the output SNR of the response of the underdamped harmonic-Gaussian double-well potential system in Eq. (17) is derived as

$$\text{SNR} = \frac{\pi c^2 \alpha_1^2 \eta_0^2}{\alpha_0^2 + \Omega^2}\Big|_{\Omega=\omega_0} \times \frac{\alpha_0^2 + \omega_0^2}{4c^2 \alpha_0}\left[1 - \frac{\alpha_1^2 \eta_0^2}{2(\alpha_0^2 + \omega_0^2)}\right]^{-1} = \frac{\pi \alpha_1 \eta_0^2}{4}\left[1 - \frac{\alpha_1^2 \eta_0^2}{2(\alpha_0^2 + \omega_0^2)}\right]^{-1} \quad (31)$$

where

$$\alpha_1 = \alpha_0 = \frac{\sqrt{\beta_1 \beta_2}}{\pi}\sqrt{-\frac{\lambda_1}{\lambda_2}}\exp(-u) \quad (32)$$

Figures 4(a)-4(d) show the output SNR as noise intensity $D$ varies under different system parameters. It is found from Fig. 4(a) that the output SNR increases and then decreases as noise intensity increases, suggesting that a noise-induced underdamped harmonic-Gaussian double-well potential SR happens. Moreover, increasing $k$ can maximize the output SNR. Like this, adjusting $\gamma$, $\alpha$ and $\beta$ can also improve the output SNR as shown in Fig. 4(b), Fig. 4(c) and Fig. 4(d) respectively, where the peak value of output SNR and the resonant noise intensity are changed. Figures 4(e)-4(h) show the output SNR as the function of system parameters for a given signal. Adjusting the system parameters can activate the underdamped harmonic-Gaussian double-well potential SR, as shown in Fig. 4(e)-4(h). Different from the overdamped harmonic-Gaussian double-well potential SR, it is noticed from Fig. 4(e)-4(h) that the antiresonance disappears in the underdamped one. That is because the damped factor changes the stability of the nonlinear system.

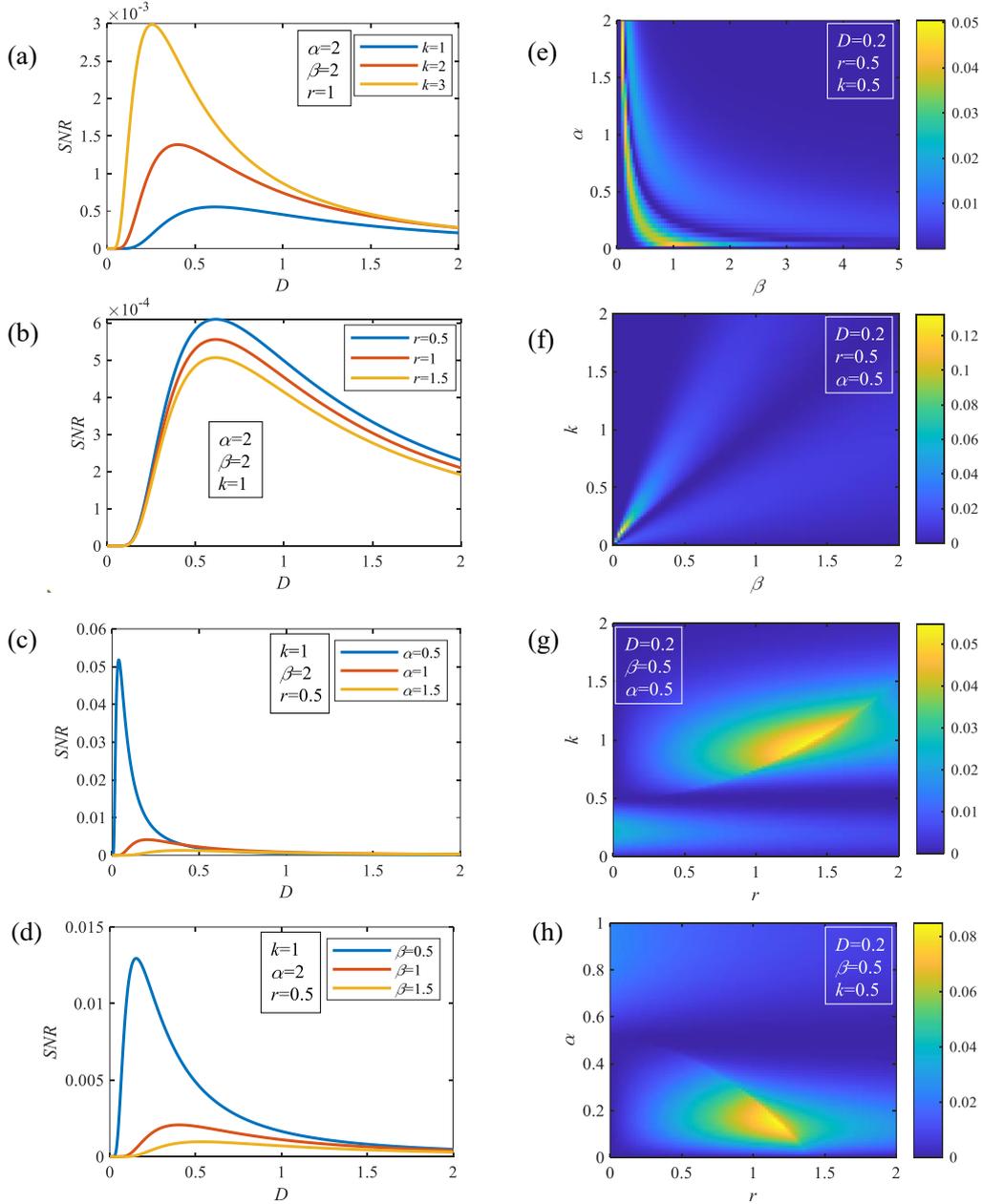

**Fig. 4** SNR of underdamped harmonic-Gaussian double-well potential SR varies with system parameters and noise intensity: SNR as a function of noise intensity under different $k$ in (a), $\gamma$ in (b) and $\alpha$ in (c), $\beta$ in (d); SNR as a two-dimensional function of $(\beta, \alpha)$ in (e), $(\beta, k)$ in (f), $(\gamma, k)$ in (g) and $(\gamma, \alpha)$ in (h).

Figure 5 shows the SPD functions and the corresponding system responses. In Fig. 5(a), the SPD functions vary from asymmetrical peaks into two symmetrical ones as noise intensity raises, suggesting that the underdamped harmonic-Gaussian double-well potential SR changes from intra-well SR into inter-well one. In Fig. 5(b), a weak period occurs when intra-well SR happens, and then the system response becomes chaotic when the particles jump randomly between double wells and finally is periodic when the inter-well SR takes place.

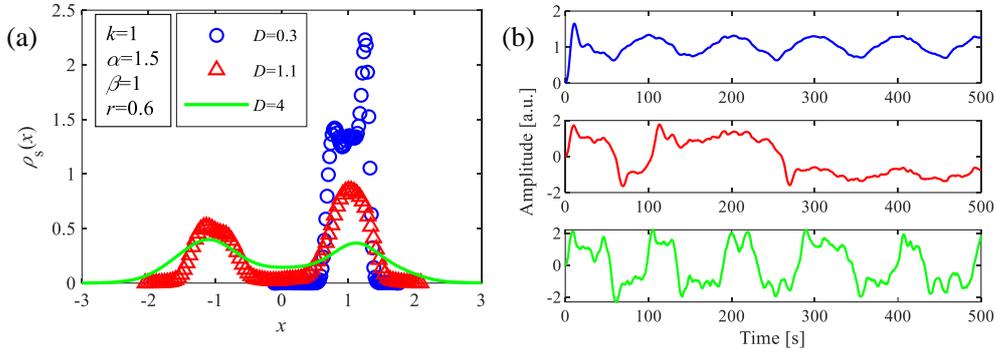

**Fig. 5** SPD functions and the corresponding system responses of underdamped harmonic-Gaussian double-well potential SR under different noise intensity: (a) the SPD functions and (b) the corresponding system responses.

## 4. Application of harmonic-Gaussian double-well potential SR to enhance weak fault characteristics of machinery

Rotating components of machinery including bearings, gears and rotors are more prone to failures than fixed components due to contact fatigue, uneven lubrication, misalignment and so on [36-38]. Therefore, how to detect weak fault characteristics of rotating components in the early stage becomes a challenge [39]. Lots of scholars have attempted to cancel or suppress the noise embedded in a signal to extract weak fault characteristics further [40, 41]. On the contrary, we would apply harmonic-Gaussian double-well potential SR to enhance weak fault characteristics of machinery by using noise.

Four Rexnord ZA-2115 double row bearing run-to-failure experiments under the rotating speed 2000 rpm and radial load 6000 lbs were performed to acquire the bearing failure data by using accelerometers and a data acquisition card. The bearing parameters are listed as below: the ball number 16, the pitch diameter 2.815 inches, the contact angle 15.17 degrees and rolling element diameter 0.331 inches. The bearing experimental rig is shown in Fig. 6(a) and the corresponding sensor placement is illustrated in Fig. 6(b). This experimental rig is composed of four tested bearings, an AC motor and rub belts [42]. In the bearing run-to-failure experiment, the sampling frequency is 20 kHz and the sampling time is 1.024 seconds.

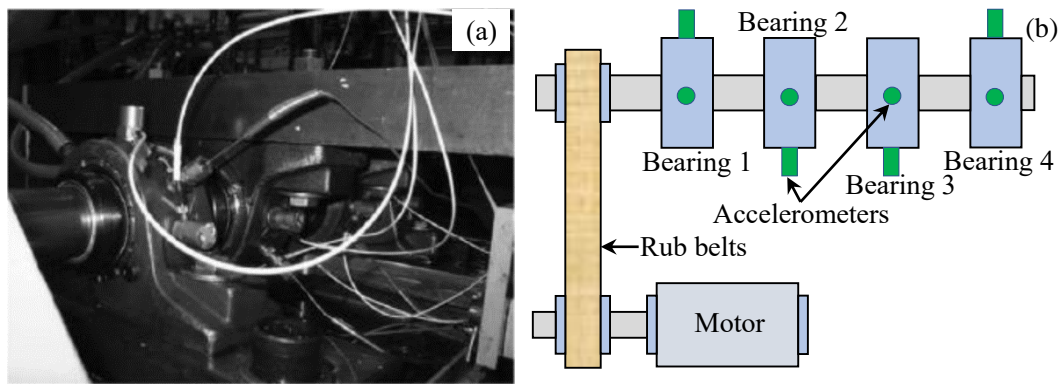

**Fig. 6** Bearing test rigs and sensor placement illustration: (a) bearing test rigs and (b) sensor placement illustration.

All failures occurred after exceeding designed life time of the bearing which is more than 100 million revolutions. The data set describes a test-to-failure experiment and consists of individual files that are 1.024-seconds vibration signal snapshots recorded at 10-minutes intervals. The recording duration is from February 12, 2004 10:32:39 to February 19, 2004 06:22:39. At the end of the test-to-failure experiment, outer race failure occurred in the tested bearing 1. The root mean square (RMS), an effective health indicator, is often used to reflect the vibration intensity and monitor the health state of bearings further. Therefore, RMS of bearing run-to-failure experimental vibration data is calculated and depicted in Fig. 7 to observe the degradation trend of the tested bearing 1. The degradation trend changes slowly with slight fluctuation in the range of 0~88 hours and then raises into the larger RMS for degradation marked in red dot in the zoomed RMS plot, suggesting that a tiny outer race failure occurs in the early stage of the tested bearing 1. As time went on, it can be seen from Fig. 7 that RMS keeps increasing, indicating that the outer race failure becomes more and more severe. Finally, this test-to-failure experiment was stopped because of strong vibration. In the test-to-failure experiment, the health state of the tested bearing varies from normal to the early failure to severe failure to end of life, which is consistent with the degradation trend reflected by RMS.

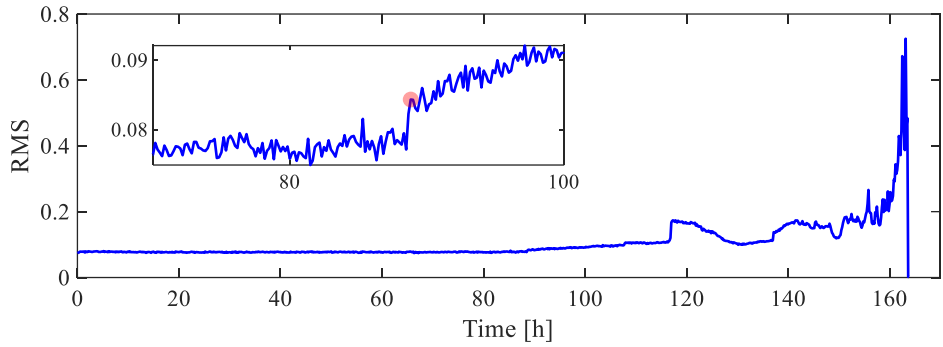

**Fig. 7** RMS of the bearing run-to-failure vibration signals.

The raw vibration signal at the 88.83*th* hour marked in red dot and its frequency and envelope spectrum are depicted in Fig. 8. We cannot observe the eye-catching spectral peaks at the theoretical outer race/inner race/roller/cage fault characteristic frequency and its harmonics from both the frequency spectrum in Fig. 8(b) and the zoomed envelope spectrum in Fig. 8(c), which are submerged by other spectral peaks from background noise and excited by other normal components. Although we have completed the bearing run-to-failure experiment and observed that a failure occurred at the outer race of the tested bearing 1 by disassembling four tested bearings, we cannot judge what time a tiny failure occurs at the outer race of the tested bearing 1 by virtue of the raw vibration signal and its spectrum in Fig. 8, which is very important for early fault diagnosis and remaining useful life prediction.

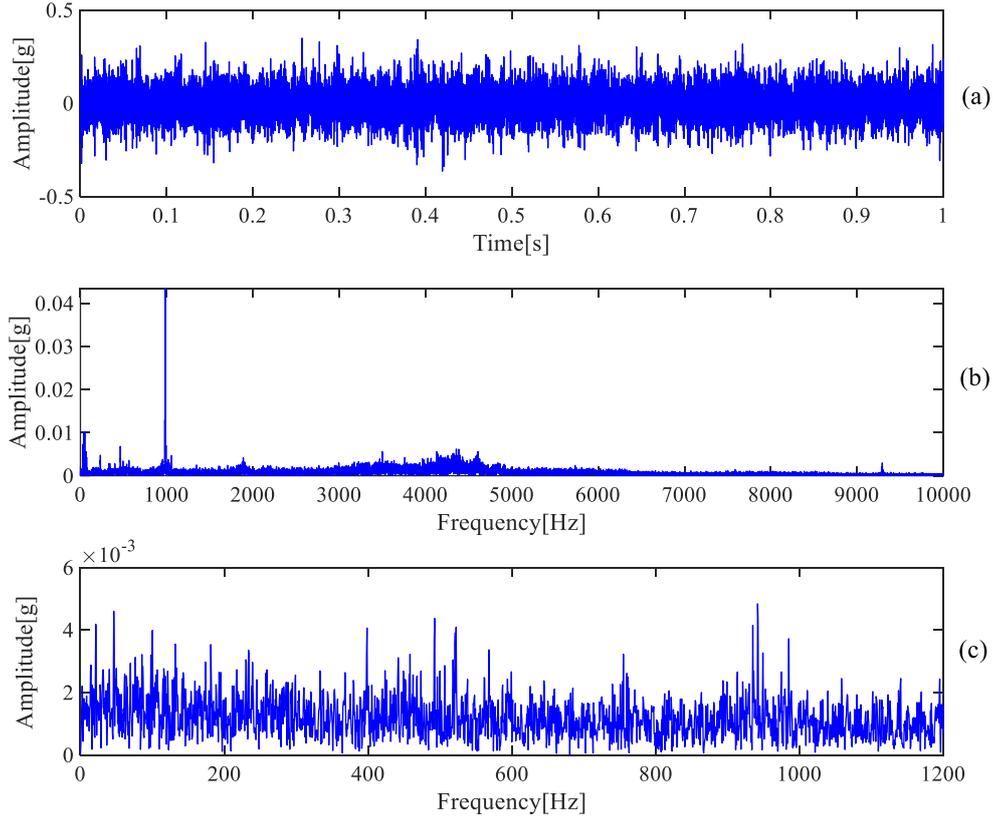

**Fig. 8** The vibration signal and its spectrum of outer race failure bearing at the early stage: (a) the raw signal, (b) its frequency spectrum and (c) zoomed envelope spectrum.

We apply the overdamped harmonic-Gaussian double-well potential SR to enhance the weak fault characteristics in the early stage of the tested bearing 1. Figure 9 shows the enhanced results of weak fault characteristics embedded in the raw vibration signal, where the system parameters are given as $k = 1.1$, $\alpha = 1.2$, $\beta = 0.24$ and the integral step is $h=0.035$. The overdamped SR cannot be used to process large-parameter signals directly and frequency-shifted and rescaling transform is widely to solve it. Three key parameters of frequency-shifted and rescaling transform in the overdamped harmonic-Gaussian double-well potential SR are given as below by virtue of the theoretical outer race fault characteristic frequency 236.4 Hz that can be calculated according to the structural parameters and rotating speed of the tested bearing 1: the pass-band cut-off frequency 220 Hz, the stop-band cut-off frequency 200 Hz and the carrier frequency 200 Hz. These parameters in the frequency-shifted and rescaling transform could be selected according to the reference [43]. One can observe from Fig. 9 that the enhanced signal characterizes strong impacts and dominant spectral peaks are at the outer race fault characteristic frequency and its second harmonic of the tested bearing 1, suggesting that a tiny failure occurs at the outer race of the tested bearing 1. However, the overdamped harmonic-Gaussian double-well potential SR depends on the high-pass filter to perform the frequency-shifted and rescaling transform, whose parameters are given artificially.

In the overdamped harmonic-Gaussian double-well potential SR-based enhanced results, the low-frequency components of the raw vibration signal (<200Hz) have been removed by using the frequency-shifted and rescaling transform. Moreover, the overdamped harmonic-Gaussian double-well potential SR method would suppress the components beyond the nonlinear filtering frequency band of overdamped harmonic-Gaussian double-well potential SR. Although overdamped harmonic-Gaussian double-well potential SR method is able to utilize the noise located in the nonlinear filtering frequency band of overdamped harmonic-Gaussian double-well potential SR for enhancing weak fault characteristics, a part of noise is removed. Therefore, the amplitude of the detected result in Fig. 9 is smaller than that in Fig. 8.

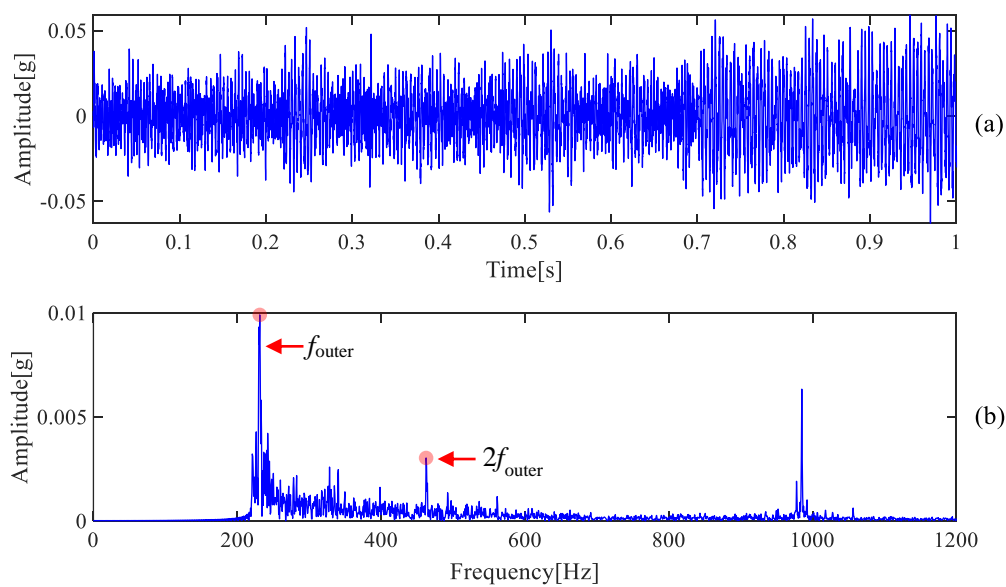

**Fig. 9** Overdamped harmonic-Gaussian double-well potential SR-based enhanced results: (a) the enhanced signal and (b) its zoomed frequency spectrum.

Further, we apply the underdamped harmonic-Gaussian double-well potential SR to enhance weak fault characteristics embedded in the raw vibration signal, as shown in Fig. 10 whose system parameters are given as $k=1.2$, $\alpha=1.1$, $\beta=0.24$, $\gamma=0.33$ and $h=0.035$. There are obvious repetitive transients in the enhanced signal and eye-catching spectral peaks at the outer race fault characteristic frequency and its second harmonic in the zoomed frequency spectrum as shown in Fig. 10(b). Compared with the overdamped harmonic-Gaussian double-well potential SR-based results, the underdamped one characterizes the higher spectral peaks at the outer race fault characteristic frequency and its second harmonic in the zoomed frequency spectrum.

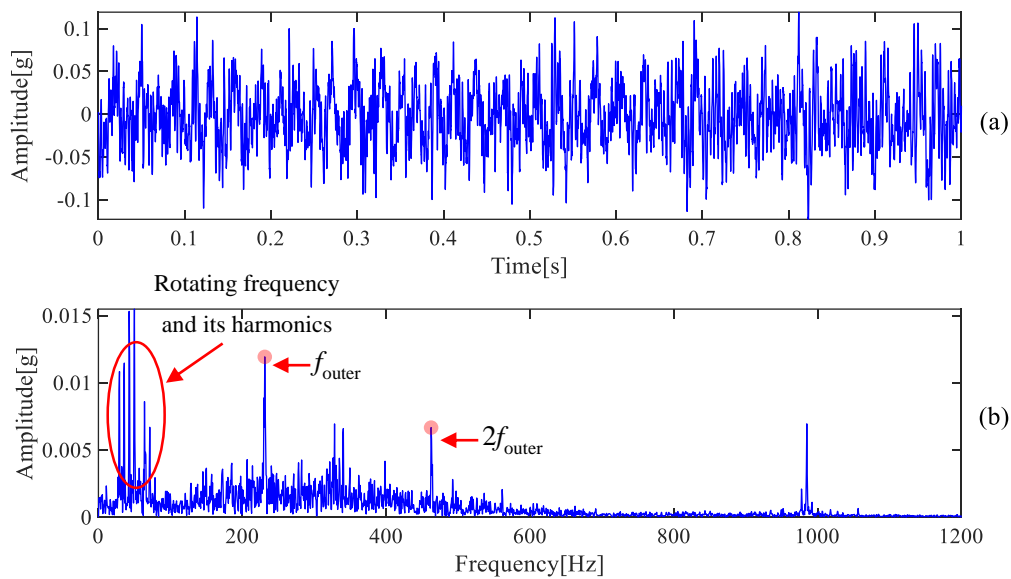

**Fig. 10** Underdamped harmonic-Gaussian double-well potential SR-based enhanced results: (a) the enhanced signal and (b) its zoomed frequency spectrum.

For a comparison, we use the advanced robust local mean decomposition (RLMD) [44, 45] to decompose the raw vibration signal of the tested bearing 1 into the product functions (PFs) and a residual component (Res) for extracting weak fault characteristics. The product functions and their zoomed envelope spectrum are shown in Fig. 11(a) and Fig. 11(b), respectively. One cannot observe the obvious spectral peaks at the outer race fault characteristic frequency and its harmonics from the zoomed envelope spectrum.

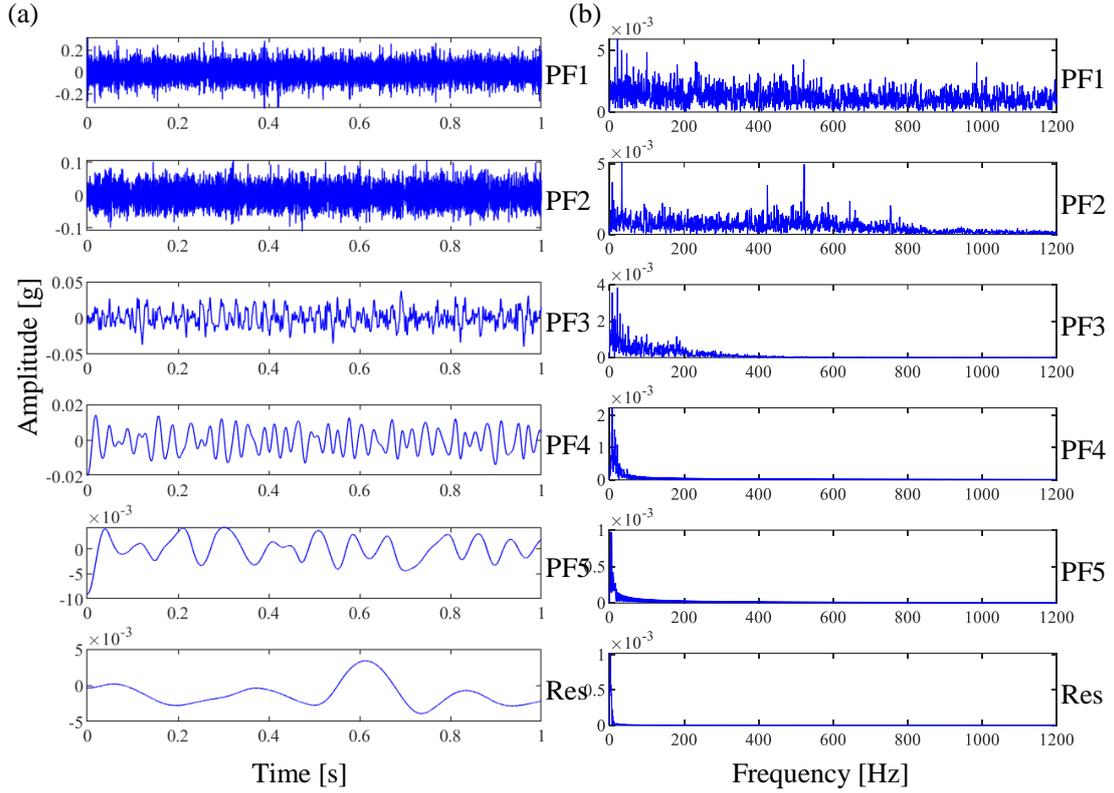

**Fig. 11** RLMD-based results: (a) product functions and (b) their zoomed envelope spectrum.

In addition to signal decomposition methods, signal denoising or signal filtering methods also have been widely applied to extract weak fault characteristics of machinery. Among them, wavelet transform [46, 47] is typical to obtain a denoised version of the raw signal by thresholding the wavelet coefficients. Here, the maximal overlap discrete wavelet transform is used to denoise the signal with soft thresholding, level 3 and db4 wavelet. The denoised signal and its zoomed envelope spectrum are shown in Fig. 12 and Fig. 13, respectively. It is found from Fig. 12 that the wavelet transform can cancel strong background noise, but we cannot see any fault characteristics at the first sight from the zoomed envelope spectrum in Fig. 13.

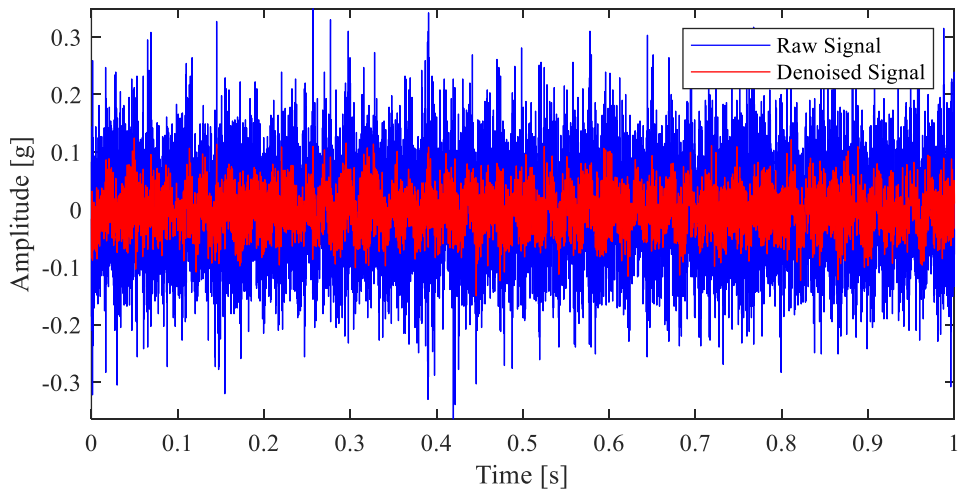

**Fig. 12** Undecimated wavelet transform-based denoised signals.

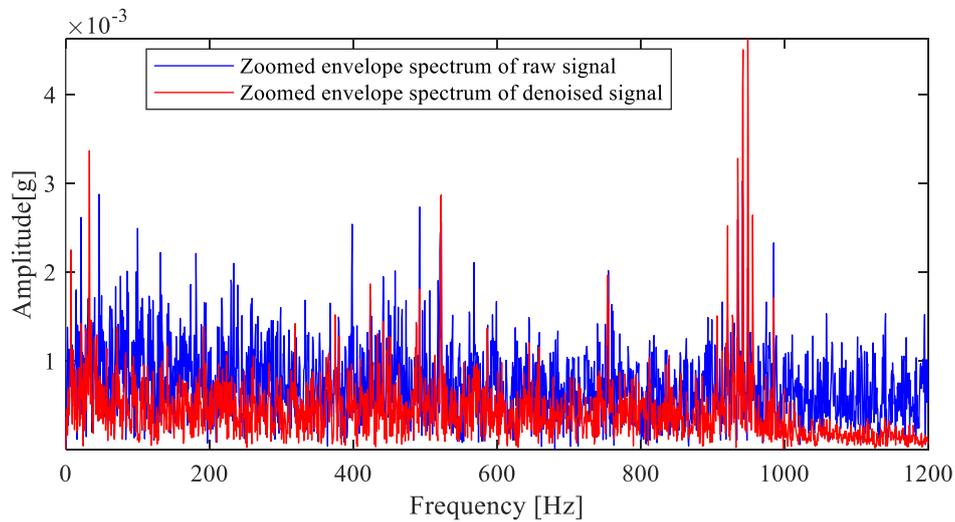

**Fig. 13** The zoomed envelope spectrum of undecimated wavelet transform-based denoised signals.

A classical symptom of rotating machines failures in vibration signals is the presence of repetitive transients. Antoni [48] proposed an infogram method to capture the signature of repetitive transients in time domain, which is the variant of classical fast kurtogram method. This method is used to process the raw vibration signal for extracting repetitive transients in time domain. The corresponding results are shown in Fig. 14. Although it can see the slight repetitive transients from the filtered signal in Fig. 14(b), it is difficult for us to identify the period of repetitive transients because of strong background noise and other normal vibration components. The above conclusion could be further confirmed by the squared envelope amplitude sepctrum of the filtered signal in Fig. 14(b), in which we cannot see the eye-catching spectral peaks at the outer race fault characteristic frequency and its harmonics.

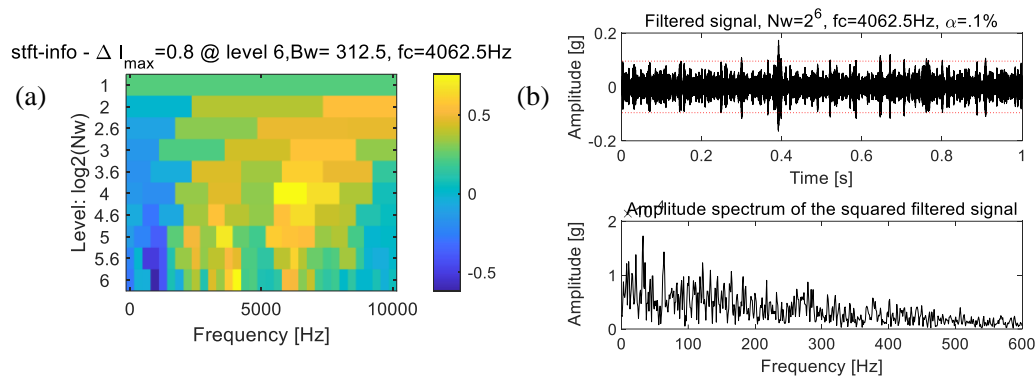

**Fig. 14** The detected results using infogram: (a) infogram and (b) the filtered signal and its squared envelope amplitude sepctrum.

## 5. Conclusions

The overdamped and underdamped harmonic-Gaussian double-well potential SR are investigated by deriving the output SNR and SPD functions. It is found that both noise-induced SR and parameter-induced SR can be activated in the overdamped and underdamped harmonic-Gaussian double-well potential systems. Moreover, since the harmonic-Gaussian double-well potential in the range of $k \geq 2\alpha\beta$ loses the stability, we can observe the antiresonance, whereas adding the damped factor into the overdamped harmonic-Gaussian double-well potential system can change the stability, resulting that the antiresonance disappears. Finally, we apply both the overdamped and underdamped harmonic-Gaussian double-well potential SR to enhance weak fault characteristics of bearings for incipient fault identification. The weak fault characteristics are enhanced successfully to identify the early failure of bearings, which somewhat outperforms to the RLMD, wavelet transform and infogram-based results. But the SR-based methods depend on the prior knowledge of the signals to be detected or structural parameters and rotating speeds of bearings, and cannot detect unknown multiple-frequency and multiple-component coupled signals without any prior knowledge. Therefore, we would study the SR-based signal decomposition method by using noise to decouple and detect unknown multiple-frequency and multiple-component signals, especially time-varying nonstationary signals in the future.

**Acknowledgments**

This research was supported by Foundation of the State Key Laboratory of Performance Monitoring and Protecting of Rail Transit Infrastructure of East China Jiaotong University (HJGZ2021114), Foundation of Yangjiang Research Center Advanced Energy Science and Technology Guangdong Laboratory (Yangjiang Offshore Wind Power Laboratory) (YOWPL2022005), Zhejiang Provincial Natural Science Foundation of China (LQ22E050003), General Scientific Research Project of Educational Committee of Zhejiang Province (Y202043287), Projects in Science and Technique Plans of Ningbo City (2020Z110), Ningbo Science and Technology Major Project (2022Z057, 2022Z002), National Natural Science Foundation of China (51905349, 62001210, U2013603), Natural Science Foundation of Guangdong Province (2022A1515010126, 2020A1515011509), Ningbo Natural Science Foundation (2022J098) and also sponsored by K.C. Wong Magna Fund in Ningbo University. The Spanish State Research Agency (AEI) and the European Regional Development Fund (ERDF) under Project No. PID2019-105554GB-I00 is also aknowledged.

**Conflicct of Interest**

The authors declare that they have no conflict of interest.

**Data availability**

The datasets generated during and/or analysed during the current study are available from the corresponding author on reasonable request.